\documentclass{article}[12pt]
\begin{document}
{\noindent \small{submitted for publication (January 2000)}}
\vskip 2cm
\begin{center}
{\Large \bf Non-gaussian electrical fluctuations \\ 
in a quasi-2d packing of metallic beads}

\vskip 1cm
{\large \bf N.Vandewalle\footnote{corresponding author, e-mail:
nvandewalle@ulg.ac.be}, C.Lenaerts and S.Dorbolo,}

\vskip 1cm
GRASP, Institut de Physique B5, Universit\'e de Li\`ege, B-4000 Li\`ege,
Belgium.

\end{center}


{\noindent keywords: granular media, electrical resistance, mechanical stresses}

\vskip 1.5cm
{\noindent \large Abstract}
\vskip 0.5cm
The electrical properties of a two-dimensional packing of metallic beads are studied. Small mechanical perturbations of the packing leads to giant electrical fluctuations. Fluctuations are found to be non-gaussian and seem to belong to L\'evy stable distributions. Anticorrelations have been also found for the sign of these fluctuations. 

\newpage

The granular state of matter exhibits fascinating physical properties such as memory effects \cite{vanel}, phase segregation \cite{segr},
heterogeneous mechanical stresses \cite{stress}, etc... Although many
experiments have been performed for studying mechanical and geometrical
aspects of the granular matter \cite{stress}, only very few reports \cite{branly,bideau,roux} can be found in the scientific literature for describing electrical properties of such systems. In 1890, Branly reported \cite{branly} that the electrical resistance of
packed metallic grains is affected by an incident electromagnetic wave. The Branly's coherer was the first reported antenna. He discovered also that a small vibration of the packing affects the electrical resistance. More recently, Ammi et al. \cite{bideau} measured a non-Hertzian behavior of the electrical conductivity when a packing is submitted to high uniaxial pressures. In an other experiment, Vandembrouck et al. \cite{roux} visualized the electrical paths in a packing of metallic beads for different injected current densities. 

In the present letter, we report electrical measurements of a 2d packing of metallic beads submitted to small perturbations. Electrical fluctuations are statistically analyzed. A physical interpretation is given within the framework of Hertzian contacts. Comparison with recent experiments and models for stress fluctuations are also emphasized. 

Our experimental set-up is illustrated in Figure 1. The main part of the
system is a tilted plane of epoxy of $25 \times 30$ $cm^2$ on which $PbO_2$ beads have been placed. A small tilt angle (typically $\theta=10^{o}$) can be adjusted by two screws. This plane has been made using printed circuit technique allowing to design easily the configuration of the electrical contacts. In our experiments, three contacts (A, B, C) have been used as described in Figure 1. Contacts are placed on specific beads. A plexiglas protection plane has been placed just above the epoxy plane to ensure the quasi-2d characteristic of the system. The mean diameter of the beads is 2.35 mm and a polydispersity of 2 \% has been measured. About 3000 beads have been placed on the epoxy plane. Above the protection plane, a wooden hammer driven by the parallel port of the computer can hit the epoxy plane at the bottom of the packing.

The experiments have been realized in a constant current regime. A current source injects a current $i$ through A and C. The tension $U$ between the point B and C is then sent to the computer by a nanovoltmeter. This method is the so-called ``3-points measurement". In each series of measurements, the following procedure has been repeated a large number $n$ of times: (i) the beads are placed in between the epoxy and plexy planes, (ii) the hammer hits the bottom of the packing every 10 seconds, (iii) the computer waits 3 seconds and then stores the tension between B and C at sampling rate of 0.2 seconds during the next 7 seconds.

Figure 2 presents a typical evolution of the voltage $U$ as a function of the shock number $n$. The inset illustrates the tension at the scale of a few seconds: shocks events are denoted by arrows at that scale. When a shock occurs, $U$ is seen to be drastically affected and thereafter sticks at that new value. One observes also that there is apparently no systematic behavior at each shock: (i) the voltage fluctuation is either positif or negatif, and (ii) the jumps/drops of $U$ seem to be characterized by a broad distribution of amplitudes. A compaction of the packing is also observed at the early stages of the experiment, and therefore one observes that the voltage $U$ globally decreases since more and more beads are touching. After $n=2000$ shocks, we have observed no macroscopic drift of the electrical voltage in our measurements! After $t=10000$ shocks, the packing is still disordered. Indeed, numerous departures (defects) from an ordered structures are observed.

Fluctuations $\Delta U_n$ are defined as differences of the electrical voltage before $U_n$ and after $U_{n+1}$ the shocks. Since $i$ is constant during the experiment, fluctuations $\Delta U$ are mainly due to changes of the conductivity of the packing, i.e.
\begin{equation}
\Delta U = i \Delta R \approx - i {\Delta \sigma \over \sigma^2}
\end{equation} from the Ohm's law $U=Ri$. From Eq.(1), one expects that the mean size of the fluctuations $\langle \Delta U \rangle$ would be proportional to the injected current $i$. In order to compare our different data series, we normalized $\Delta U$ by $i$. Figure 3 presents the distribution of fluctuations $P(\Delta R)$ in a semi-log plot for different values of the injected current $i$. So-called ``fat tails" for rare events are observed in Figure 3, the ocurrence of large fluctuations is several orders of magnitude more frequent than those for gaussian tails. The fluctuation histograms are however well described by a L\'evy stable distribution which has the following behaviors:
\begin{equation}
P(|\Delta R|) \sim \exp{\left( - \gamma |\Delta R|^{\alpha} \right)} \hskip 0.5cm for \hskip 0.5cm |\Delta R| < < 1/\gamma
\end{equation} and
\begin{equation} 
P(|\Delta R|) \sim |\Delta R|^{-(\alpha +1)} \hskip 1cm for \hskip 0.5cm |\Delta R| > > 1/\gamma
\end{equation} where $0 \le \alpha \le 2$ is the so-called L\'evy index and $\gamma > 0$ is a quantity related to the width of the distribution (for $\alpha > 1$). When $\alpha$ reaches 2, the distribution becomes a Gaussian distribution. When $\alpha = 1$, the L\'evy distribution reduces to a Cauchy distribution. When $\alpha < 1$, the second moment of the distribution diverges. The continuous curves in Figure 3 are fits with Eq.(2) giving a value $\alpha=0.9 \pm 0.1$ for both cases. Figure 4 presents in a log-log plot the fat tail of the distribution $P(|\Delta R|)$. Power law behaviors (Eq.(3)) are observed. We have found $\alpha = 0.95 \pm 0.10$ in good agreement with the $\alpha$ value of the first fits using Eq.(2). Both fits in Figures 3 and 4 confirm the L\'evy scenario for electrical fluctuations. This value is quite low (lower than a Cauchy) and expresses that the voltage evolves erratically with frequent large drop and burst excursions. It is also worthwile to point out that the L\'evy index $\alpha$ seem to be independent of the injected current $i$ ranging from 3$mA$ up to 100$mA$ in our different series of measurements.  

Let us also investigate the temporal correlations between two successive fluctuations. Figure 5 presents the phase space $(\Delta U_n , \Delta U_{n+1})$ for a current $i$=50$mA$. The data points are mainly dispersed in an ellipsoid shape which is tilted with respect to the horizontal and vertical axis. Moreover, branches are observed along the vertical, horizontal and diagonal directions. The latter branches indicate the existence of strong correlations for large fluctuations. The continuous line is a linear fit through the entire cloud of data points. The slope $a=-0.45 \pm 0.01$ of the linear regression $\Delta U_{n+1} = a \Delta U_n + b$ is significantly negative, and expresses that two successive fluctuations are highly anticorrelated. The slope $a$ is found to be weakly depend on $i$.

Finally, it should be noted that $Ni$ beads which are more conducting than $PbO_2$ beads lead to similar results: L\'evy-like distributions and anticorrelations. We checked also that the results are not sensitive to the tilt angle $\theta$.

If the properties of the fluctuations characterizing the phenomenon are found to obey non-trivial laws, they should certainly arise from fundamental phenomena. A physical interpretation of these giant and non-Gaussian fluctuations should be found in the framework of the geometry of electrical paths. Indeed, the existence of multiple tortuous conducting paths have been reported in \cite{roux}. We conjecture that the fluctuations observed herein are most probably related to the fluctuations of stress chains in the packing. Also, the strong anticorrelations that we put into evidence should be related to memory effects in the packing.

As recently underlined by deGennes \cite{degennes}, both macroscopic electrical conducitivity $\sigma$ and macroscopic elastical modulus $\mu$ of the packing are closely related if Hertzian contacts are assumed. This analogy between mechanical strain and electrical conductance implies that the distribution of the fluctuations measured herein could be related to the stress fluctuations in the packing. Giant stress fluctuations have been recently reported in compression \cite{rajchenbach} and photoelastic experiments \cite{howell}. Cooperative rearrangements such as nucleation of cracks and the propagation of these cracks in the packing can lead to stress flcutuation of high amplitude. It has been put into evidence that theses cracks have a transient life and heal after the displacement of blocks or single beads \cite{rajchenbach}. From the simulation point of view, Coppersmith et al. \cite{qmodel} introduced the so-called $q$-model deducing a probability distribution of vertical forces varying as $f \exp{(-f/ \langle f \rangle)}$. Howerer, the $q$-model and more elaborated variants lead to Gaussian distributions of stress fluctuations. At our knowledge, the only one model which is able to produce rearrangements of force paths at all scales is the so-called Scalar Arching Model (SAM) introduced by Claudin et al. \cite{claudin}. The SAM gives rise to a broad distribution of the apparent weight variations $\Delta W_a$ measured at the bottom of a silo. Claudin has found \cite{claudin2} a universal power law behavior $P(|\Delta W_a|) \sim |\Delta W_a|^{-0.94}$ whatever the values of solid friction coefficients. This power law behavior is consistent with our results but the relationship between our exponent $\alpha+1$ and the Claudin's exponent 0.94 is non-obvious. 

In summary, the careful measurements of the electrical properties of a quasi-bidimensional packing of $PbO_2$ beads have put into evidence the non-gaussian character of conductivity fluctuations. Those non-Gaussian electrical fluctuations are related to stress fluctuations in the packing. Our measurements confirm some predictions of the Scalar Arching Model.

\vskip 1.5cm
{\noindent \large Acknowledgements}
\vskip 0.4cm

NV thanks the FNRS (Brussels, Belgium). SD is financially supported by FRIA (Brussels, Belgium). Valuable discussions with M.Ausloos, E.Clement, R.Cloots, J.Rajchenbach, S.Galam, P.Harmeling and Ph.Vanderbemden are acknowledged.

\vskip 1.5cm
{\noindent \large Figure Captions}
\vskip 0.5cm

{\noindent \bf Figure 1} -- A schematic illustration of the experimental setup. Contacts A, B and C on beads. Epoxy (E) and Plexiglass (P) Planes which are separated by the bead diameter and which are tilted by an angle $\theta$.

\vskip 0.5cm
{\noindent \bf Figure 2} -- The electrical fluctuations of $U(n)$ in the quasi-2d packing of $PbO_2$ beads. The inset presents the evolution of $U$ at the scale of a few seconds.

\vskip 0.5cm
{\noindent \bf Figure 3} -- Semi-log plot of fluctuation distributions $P(\Delta R)$ for two different current intensities $i$. The continuous curves are fitted L\'evy-like distributions using Eq.(2).

\vskip 0.5cm
{\noindent \bf Figure 4} -- Log-log plot for the tail of $P(|\Delta R|)$. The lines are power law fits using Eq.(3).

\vskip 0.5cm
{\noindent \bf Figure 5} -- Phase space $(\Delta U_n , \Delta U_{n+1})$. The continuous line is a linear regression $\Delta U_{n+1} = a \Delta U_n + b$.

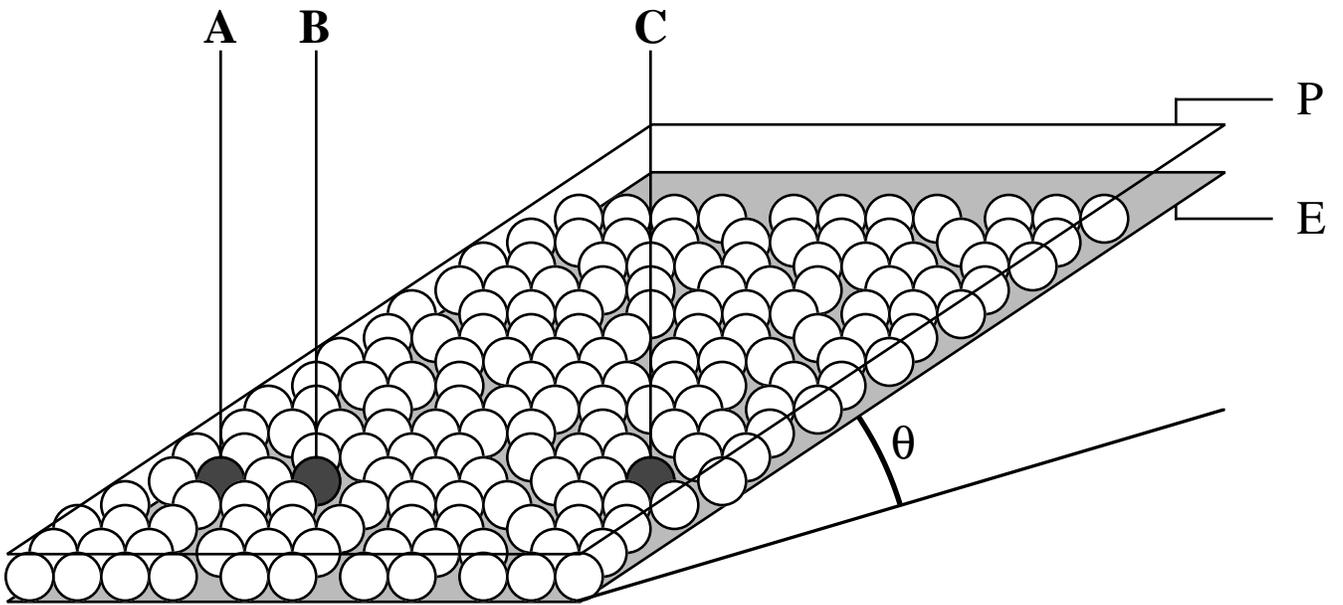
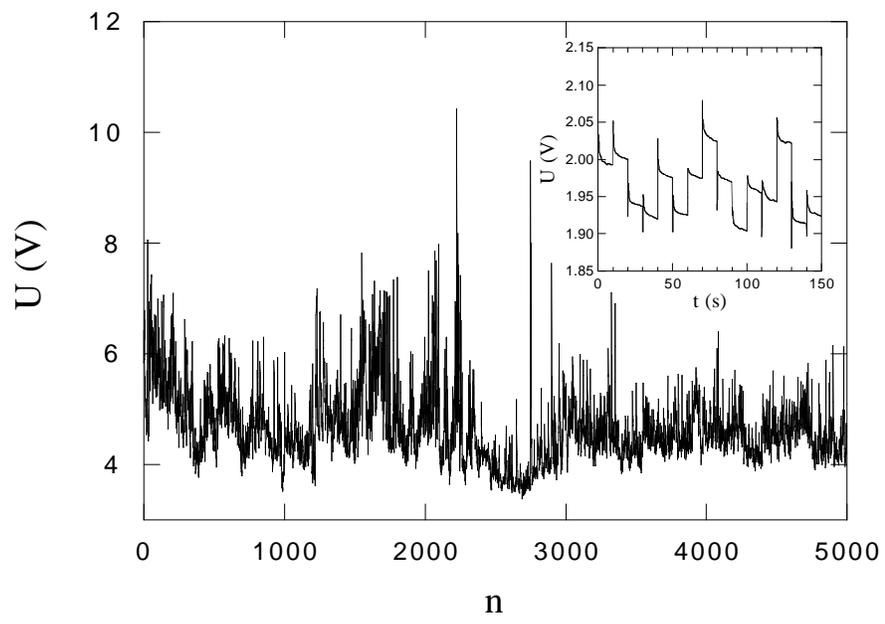
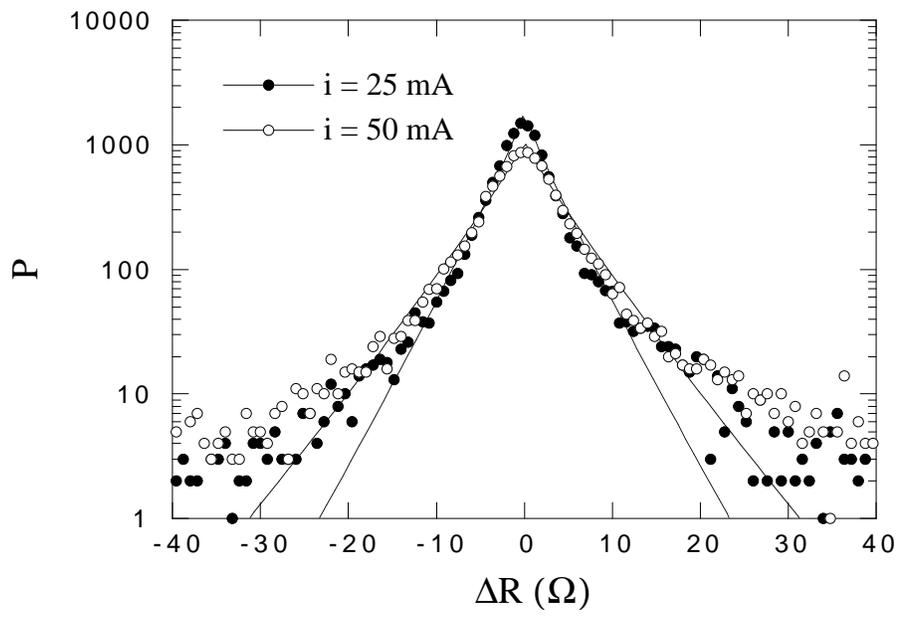
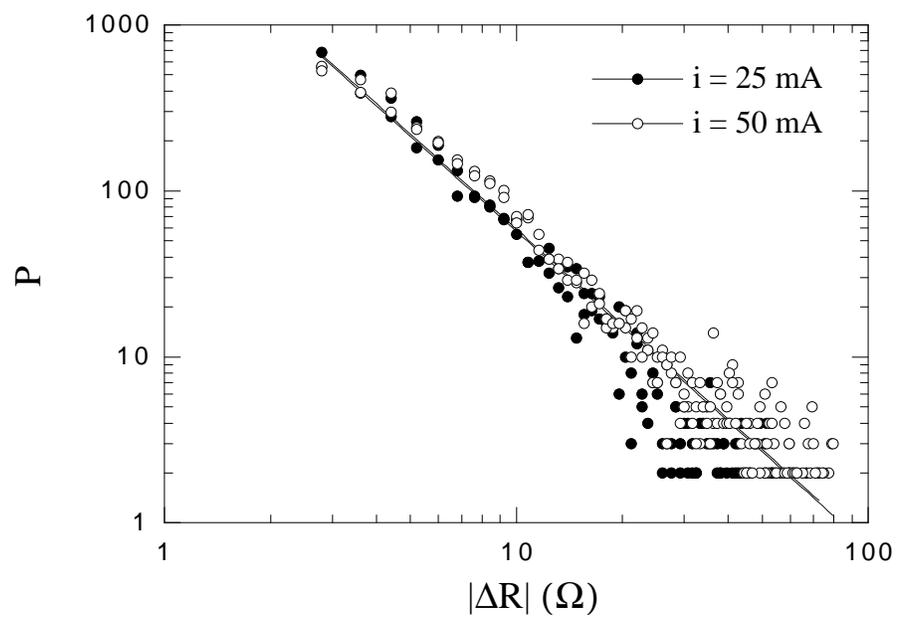
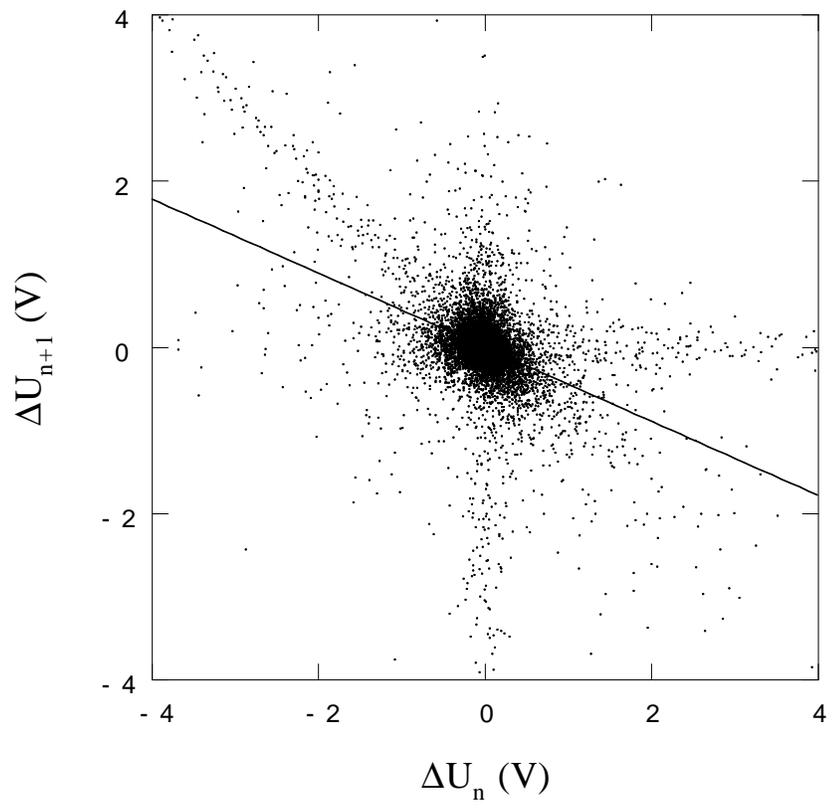
\end{document}